\documentclass[sigconf]{acmart}
\acmConference[ESEC/FSE 2018]{12th Joint Meeting of the European Software Engineering Conference and the ACM SIGSOFT Symposium on the Foundations of Software Engineering}{4--9 Nov., 2018}{Lake Buena Vista, Florida, USA}

\usepackage{booktabs} 
\usepackage{graphicx}
\usepackage{tcolorbox}

\makeatletter
\newcommand{\verbatimfont}[1]{\renewcommand{\verbatim@font}{\ttfamily#1}}
\makeatother

\setcopyright{rightsretained}

\acmDOI{10.475/123_4}

\acmISBN{123-4567-24-567/08/06}

\acmYear{2018}
\copyrightyear{2018}

\acmPrice{15.00}

\begin{document}

\title[Detecting Speech Act Types During Developer Q/A]{Detecting Speech Act Types in Developer Question/Answer Conversations During Bug Repair}


\author{Andrew Wood}
\affiliation{%
  \institution{University of Notre Dame}
  \city{Notre Dame} 
  \state{Indiana} 
  \postcode{46556}
}
\email{awood7@nd.edu}
\author{Paige Rodeghero}
\affiliation{%
	\institution{Clemson University}
	\city{Clemson}
	\state{South Carolina}
	\postcode{29634}
}
\email{prodegh@clemson.edu}
\author{Ameer Armaly}
\affiliation{%
	\institution{Google}
	\city{Mountain View} 
	\state{California} 
	\postcode{94043}
}
\email{aarmaly@nd.edu}
\author{Collin McMillan}
\affiliation{%
	\institution{University of Notre Dame}
	\city{Notre Dame} 
	\state{Indiana} 
	\postcode{46556}
}
\email{cmc@nd.edu}


\begin{abstract}
This paper targets the problem of speech act detection in conversations about bug repair.  We conduct a ``Wizard of Oz'' experiment with 30 professional programmers, in which the programmers fix bugs for two hours, and use a simulated virtual assistant for help.  Then, we use an open coding manual annotation procedure to identify the speech act types in the conversations.  Finally, we train and evaluate a supervised learning algorithm to automatically detect the speech act types in the conversations.  In 30 two-hour conversations, we made 2459 annotations and uncovered 26 speech act types.  Our automated detection achieved 69\% precision and 50\% recall.  The key application of this work is to advance the state of the art for virtual assistants in software engineering.  Virtual assistant technology is growing rapidly, though applications in software engineering are behind those in other areas, largely due to a lack of relevant data and experiments.  This paper targets this problem in the area of developer Q/A conversations about bug repair.
\end{abstract}

%
%



\maketitle

\section{Introduction}
\label{sec:intro}
``Speech Acts'' are spoken or written actions meant to accomplish a task~\cite{searle1965speech, bach1979linguistic, stent2014natural}. A classic example of a speech act is `I now pronounce you husband and wife' -- the speech itself is an action with consequences~\cite{searle1980speech}.  Naturally, most speech acts in life are less impactful (`let's go to the movies' or `please tell me how to find my classroom'), though the principle is the same.  Speech acts are key components of conversations that guide the what the speakers do.

While research in sociology has studied speech acts for decades~\cite{searle1965speech, bach1979linguistic}, there has been an increase in interest due to the growth of virtual assistants.  Virtual assistants such as Cortana~\cite{Cortana}, Google Now~\cite{GoogleNow}, Siri ~\cite{Siri}, etc., try to carry on a conversation with a human, to try to serve that person's request -- asking for a restaurant recommendation, or the time of day.  And while human conversation can seem effortless at times, in fact there are several key steps that we do without even being aware~\cite{ten2007doing, ford2002language, hutchby2008conversation,nishida2007convanalysis}: we detect when speech acts occur, we comprehend the speech act as being a particular type of act (e.g., an information request, a command, a clarification), and craft an appropriate response.  We understand naturally that the type of act will depend on the context of the conversation, and that a piece of dialog may be of more than one type.  Virtual assistants must be carefully designed to mimic this process: the first step is to \emph{detect} speech acts and classify them by type.

Designing a virtual assistant to detect and classify speech acts requires examples of conversations from which to learn what those speech acts are.  These conversations must be related to the task for which the assistant is being designed.  For example, a study by Whittaker~\emph{et. al}~\cite{whittaker2002fish} targets dialog systems for restaurant recommendations, and therefore collects 24 examples of conversations in which a human asks for restaurant recommendations.  Kerly~\emph{et. al}~\cite{KERLY2007177} targets automated tutoring systems, and to do so collects 30 examples of tutoring sessions for a specific subject area.  The data collected for one application domain is generally not applicable to other domains.

One key, accepted strategy for collecting examples of conversations is a user simulation in a ``Wizard of Oz'' experiment~\cite{DAHLBACK1993258, rieser2011reinforcement}.  In a Wizard of Oz experiment, human participants interact with a machine that the participants believe to be automated.  In reality, the machine is controlled by human experimenters.  The participants are asked to use the machine for a particular purpose (e.g., they ask for a restaurant recommendation).  The idea is that the experimenters can collect simulated conversations that closely reflect real-world use.  Analysis of the conversations reveals what speech acts the participants make, and clues as to how to detect them.

Today, virtual assistants are possible due to major efforts in understanding human conversation, though these efforts have largely been confined to everyday tasks.  While virtual assistants for software engineering have been envisioned for decades~\cite{searle1965speech, bach1979linguistic}, progress is limited, largely due to three problems that we target in this paper: 1) there are very few experiments with data released of software engineering conversations, 2) the speech act types that software engineers make are not described in the relevant literature, and 3) there are no algorithms to automatically detect speech acts.

In this paper, we conduct a Wizard of Oz experiment in the context of bug repair. We then manually annotate the data from this experiment to find the speech act types and build and evaluate a detector for these speech acts in conversations.  Our target problem domain is a virtual assistant to help programmers during bug repair.  We chose bug repair because it is a common software engineering task, and because, as previous studies have shown, bug repair is a situation in which programmers are likely to ask questions~\cite{4497212, Ko:2004:DWD:985692.985712}.  We recruited 30 professional programmers to fix bugs for two hours each, while providing an interface to a Wizard of Oz simulated virtual assistant.  The programmers interacted with the simulated virtual assistant for help on the debugging task.  We then manually annotated each conversation with speech act types in an open coding procedure (see Section~\ref{sec:annotations}).  Finally, we trained a learning algorithm to detect speech acts in the user's side of the conversations, and evaluated its performance (Sections~\ref{sec:predictapproach} - \ref{sec:predictEvalResults}).

Across 30 two-hour conversations, we made 2459 annotations and discovered 26 speech act types.  Our automated speech act detection algorithm achieved an average of 69\% precision and 50\% recall.  By releasing this corpus, we contribute one of very few WoZ corpora, which are especially rare in the domain of Software Engineering~\cite{serban2015survey}.  We release all data, including conversations, annotations, and our detection algorithm source code via an online appendix (Section~\ref{sec:reproducability}), to promote reproducibility and assist future research in software engineering virtual agents.

\section{Problem, Significance, Scope}
\label{sec:problem}

The problem we target in this paper is that models of developer conversations are not described in the literature.  Certainly, strong efforts in the area of program comprehension have made inroads into our understanding of the types of information that programmers need and how programmers make sense of software problems.  However, the ``nuts and bolts'' of actual conversations among programmers are still not well-understood.

A key component of those nuts and bolts are ``speech acts'' (as defined in the previous section), and our goal is to automatically detect these speech acts in conversations.  But detection of speech acts is useful beyond pure academic interest: advancements in programmer tool support depend on improved detection of programmer intent.  Numerous software engineering tools depend on natural language interfaces, such as code search engines, navigation tools, traceability tools, and our target context of \textbf{automated virtual assistant} technology.  The situation we envision is that a programmer asks an automated virtual assistant a question in lieu of a fellow human programmer, and the virtual assistant is expected to provide an answer to that question.  A fundamental part of answering these questions is to detect the types of statements, comments, etc., that programmers make when asking and clarifying their questions.

Throughout this paper, we refer to a 2011 book by Rieser and Lemon~\cite{rieser2011reinforcement} as both motivation for and rationale behind our work.  The book provides an excellent summary of the design decisions required for building dialog systems and reflects the significant momentum in years of research on virtual agents -- one key theme is that using Wizard of Oz studies to inform data-driven dialog system construction is a highly effective strategy.  They point out that while it is possible to design a virtual assistant using manually-crafted assumptions about user behavior, the existence of annotated, simulated dialog (via a WoZ study) provides an immense boost to the flexibility and effectiveness of virtual agent design.  One benefit is from the increased knowledge scientists gain from studying the dialog, while another benefit is from the ability to use supervised and reinforcement learning algorithms to ``teach the computer'' correct behavior, even with relatively sparse data.

In this paper, we contribute the dataset, our manual annotation of the dataset, and our analysis of those annotations to the community as a foundation for building better software engineering virtual agents.  This contribution alone is significant, considering that a recent survey by Serban~\emph{et al.}~\cite{serban2015survey} found only four publicly-available WoZ datasets (more are held privately) suitable for building dialog systems -- and none related to Software Engineering.  However, we take a further step towards a working virtual agent by building a classifier to automatically label the dataset; in essence, this is a detector for speech act type using supervised learning (as chapter 7 of ~\cite{rieser2011reinforcement} highlights, supervised learning is often the first technique tried for speech act type detection, prior to resorting to more complex approaches).

Note that in our manual annotation process, we annotated the entire conversation (both ``Madeline's'' and the study participants' side).  However, during the speech act type detection, we only predict the type of speech acts from the participants' side of the conversation.  This is because during the manual annotation process, we study not only the participants, but the wizards' actions also: this is for the purpose of laying a groundwork for conversation flow analysis in future work, in addition to the academic interest presented in this paper.  But, during speech act detection, the realistic scenario is that a virtual assistant would never need to classify its own conversation, since it would already know the speech act types it generated itself.  It would only need to detect the speech act type of the human user.


\section{Background}
\label{sec:background}

This section describes four key technologies related to and underpinning our work in this paper: automated virtual assistants, conversation analysis and modeling, studies of program comprehension, and text classification.

\vspace{-0.1cm}

\subsection{Automated Virtual Assistants}

Automated virtual assistants such as Siri, Cortana, and Google Now are claiming an increasing role in computing for everyday tasks.  They simplify duties such as planning meals and finding music, and are part of a broader trend towards automated productivity services.  Virtual assistants for software engineering have been envisioned for decades~\cite{boehm2006view, robillard2014recommendation}, with the dream being a system that can mimic the answers that human teammates would give, such as a system able to generate ``On-Demand Developer Documentation'' as responses to source code queries~\cite{robillarddemand}.

We (the software engineering research community) are still far away from this dream.  Nevertheless, advancements are being made in that direction.  Recently, Bradley~\emph{et al.}~\cite{icse2018assist} built Devy, a virtual agent to help automate programmer tasks.  Devy differs from our work in that we seek to understand the structure of programmers' conversations, to build a system to help programmers learn and recall information, rather than automate tasks.  Pruski~\emph{et al.}~\cite{Pruski2015} created TiQi, a technique that answers database query questions in the form of unstructured dialog.  Ko and Myers~\cite{Ko:2010:EAW:1824760.1824761} created Whyline, which answers questions about program output.  Escobar-Avila~\emph{et al.}~\cite{Escobar-Avila:2017:TRT:3098344.3098461} answered unstructured questions by connecting questions to software engineering video tutorials.  A majority of current efforts focus on understanding unstructured software engineering data; for a more complete survey we direct readers to Arnaoudova~\emph{et al.}~\cite{7203123}.  But what holds back progress at the moment is an incomplete understanding of how programmers communicate -- it is not possible to build a tool that participates in this communication without understanding the nature of that communication.  This understanding can only be completed with conversation analysis and modeling.

\vspace{-0.1cm}

\subsection{Conversation Analysis and Modeling}

Conversation analysis and modeling is the task of extracting meaning from human written or verbal communication.  It usually involves creating a representation of a type of conversation (e.g., restaurant recommendations, or technical support calls~\cite{rieser2011reinforcement}), and then using that representation to predict the flow of the conversation.  A ``flow'' of a conversation is how people tend to put information in conversations, for example one conversation participant asking ``does that make sense?'' if the other participant is silent after receiving new information.  Conversations are typically broken up by \emph{turns}~\cite{ten2007doing, ford2002language, hutchby2008conversation, nishida2007convanalysis, emanuel2007sequence}. A turn begins every time a speaker begins speaking and can encompass multiple sentences. Conversation analysis and modeling is what allows automated virtual assistants to create human-like conversations.

Conversation modeling has its roots in sociology~\cite{ten2007doing, ford2002language, hutchby2008conversation, nishida2007convanalysis, emanuel2007sequence} and psychology~\cite{harnad1990symbol}, where researchers studied the factors behind conversation flow and form.  These often employ qualitative analysis methods to isolate human factors such as social rank or fatigue.  After a significant investment in the 1990s, quantitative analysis procedures have been developed to model and predict the types of information that human conversations include, in order to create interactive dialog systems.  Work in this area has flourished, with representative work including:~\cite{hirschman1998evaluating, Walker:2001:QQE:1073012.1073078, Reithinger:1995:USD:981658.981674, Moore:1994:PED:197236, Mittal:1995:DGF:223904.223916, D'mello:2013:AAA:2395123.2395128, Forbes-Riley:2008:RIS:1341585.1341623}.  For example, work by Lemon~\cite{lemon2011learning, rieser2011reinforcement} models restaurant recommendation conversations as a Markov Decision Process, in which each turn is one of six possible states.

A typical strategy in conversation modeling for discovering speech acts is \emph{user simulation}, in which participants in a study are told that they are interacting with a dialog system, which is actually a human acting like a dialog system via a chat program~\cite{ai2007comparing, schatzmann2007agenda}.  The simulation results in a transcript of a conversation between a human participant and an idealized virtual assistant (simulated by the researcher).  The transcript is an extremely valuable source of information on how the human participant expects to interact with a machine and how the machine should respond.  While rare in Software Engineering, these studies are not unheard of: Goodrum~\emph{et al.}~\cite{goodrum2017requirements} perform a WoZ study to discover what requirements knowledge programmers need, related conceptually to requirements-gathering WoZ studies proposed earlier~\cite{White:2003:BCL:1052829.1052854}.

\vspace{-0.15cm}

\subsection{Studies of Program Comprehension}

This paper could be broadly classified as a study in program comprehension -- how programmers comprehend and communicate about software development and behavior.  Typically questions asked by program comprehension literature relate to the mental and physical processes that developers follow~\cite{Holmes:2013:SPS:2377656.2377657, LaToza:2006:MMM:1134285.1134355}.  Examples of mental processes include targeting how code is connected~\cite{Mirghasemi:2011:QMB:2025113.2025184, Kramer:2012:BLA:2337223.2337451, Sillito:2008:AAQ:1446226.1446241, Sim:1998:ASC:580914.858229}.  Physical processes include taking of notes~\cite{ALTMANN2001189} and patterns of movements of the eyes~\cite{Rodeghero:ICSE:2014, Sharif:2012:ESR:2168556.2168642}.  Notably, Roehm~\emph{et al.}~\cite{Roehm:2012:PDC:2337223.2337254} point out that programmers ``try to avoid'' program comprehension, and look for short cuts whenever possible.  This finding is in line with several others that suggest that tool support for comprehension should provide information incrementally and at as high a level as possible, and avoid too many low-level details~\cite{5306335, Forward:2002:RSD:585058.585065, 1241364}.  Our vision in this paper is to build a foundation for software engineering virtual assistants, to provide information in the order and at the time requested by programmers during a dialog.

\vspace{-0.15cm}

\subsection{Text Classification}

Text classification is an intensely-studied area in machine learning, and text classification techniques have seen extensive use in software engineering.  A recent book by Aggarwal and Zhai~\cite{aggarwal2012mining} surveys text classification and mining techniques generally.  Software engineering applications are so prevalent that we cannot list them all here, though representative examples include~\cite{Rodeghero:2017:DUS:3097368.3097375, J:Anvik:ICSE:2006, kim2008classifying, menzies2008automated}.  We use text classification as a component of our speech act detection.

\vspace{-0.1cm}

\section{User Simulations}
\label{sec:user_simulations}

In this section, we describe our user simulation study.  In general, a user simulation is an imitation of a conversation between a human and a machine -- instead of a real machine, a researcher stands in for the machine without the human being aware of it~\cite{DAHLBACK1993258}. In this paper, our user simulation is the interaction between our participants and an imitated software program.  Participants believed the program could automatically assist programmers with tasks.  They were informed their participation in this study was helping to improve a virtual assistant program for programmers.  However, there was no actual virtual assistant producing answers to the questions asked by the participants.  We manually answered every question.  


\vspace{-0.1cm}
\subsection{Methodology}
\label{sec:user_simulation_methodology}
We based our methodology on previous studies of bug repair in software engineering~\cite{Jiang2017, gethers2012integrated, ko2007information} and previous ``Wizard of Oz'' studies in sociology~\cite{DAHLBACK1993258}.  We asked the programmers to remotely participate in the study using a provided Ubuntu 64-bit virtual machine and the Microsoft Skype application on their local machine.  We instructed the participants to fix bugs from pre-installed open source Java projects contained within an Eclipse IDE~\cite{Eclipse} workspace on the provided virtual machine.  We instructed the participants to fix as many bugs as they could within a pre-defined two-hour time frame.  During that time, we gave the participants one bug at a time, one bug per project. We asked the participants to avoid using the Internet to search for solutions or answers to any questions that they might have, and to instead direct their questions to a automated virtual assistant named ``Madeline'' through the Skype platform.  Note that this involved two key design decisions informed by Rieser and Lemon's guide on WoZ studies for dialog systems (chapter 6 of~\cite{rieser2011reinforcement}): First, we used the written text chat only, no voice, to limit the scope of the study to developer Q/A conversations instead of introducing the possibility of voice transcription errors (it is necessary to deliberately add noise to WoZ studies involving voice to simulate actual noise, and we felt this would add too many variables considering the already complicated nature of debugging).  Second, we restricted access to internet resources.  While this may seem to create an unrealistic situation (since programmers frequently use Stackoverflow, etc.), it was necessary in order to learn how programmers might use a virtual agent, due to a bias in which people might not try a new technology simply because it is unfamiliar, and to avoid biases introduced by external tools.  These restrictions are often a ``necessary evil'' in WoZ experiments -- for example, 94\% of papers surveyed by Riek~\cite{riek2012wizard} placed substantive restrictions on participant behavior and resources.

During each study, two to three of the authors collaborated at all times to respond to the participants.  At least one of the authors had previously fixed the bugs given to the participants.  This allowed for quick and intelligent responses to the participants, giving the illusion that Madeline produced responses automatically.  This deception, typical of the ``Wizard of Oz'' strategy~\cite{dahlback1993wizard} was necessary to ensure the authenticity of the responses. The participants were explicitly told that they were communicating with an automated system supervised by humans (Madeline).  The participants were told to interact with Madeline through Skype conversations, and also to share their screens for quality assurance purposes. In reality, screen sharing provided the means to prepare responses in real time and was critical for imitating a fully autonomous system.  Following Rieser and Lemon's WoZ process for dialog systems (again, chapter 6 of \cite{rieser2011reinforcement}), we did not restrict wizards to a script or set of predefined speech act types, since a goal of our study was to understand what the programmers needed rather than test a predefined script.

\subsection{Participants}
\label{sec:user_simulation_participants}

We recruited 30 professional programmers to participate in our study.  These programmers were recruited through email and an online freelance website called Upwork~\cite{UpWork}.  The programmers work at various companies such as IBM, GolfNow, and Hyland Software, while some work as freelancers full time.  Note that the programmers recruited are not students, but professionals working in industry. Each programmer recruited had familiarity with Java before participating in the study.  Overall, the participants had an average of 5.5 years of experience with Java.  The maximum number of years of Java experience was 12 and the minimum was one.   

\vfill
\vspace{-0.2cm}
\subsection{Threats to Validity}
\label{sec:user_simulation_threats_to_validity}
As with most studies, this project has a few threats to validity.  First, since each experiment was two hours long (not including any technical problems), it is possible that the participants experienced fatigue.  This is compounded with any fatigue that they already experienced from their normal work schedule.  This was mitigated by using a large pool of participants.  Another threat came from technical problems with screen sharing.  The only issue with this, however, was a possible reduction in response speed, but we saw no noticeable reductions in any of the studies.  Either through technical problems or participants forgetting to save them, a few screen shares were unable to be stored.  However, these stored recording were not actually used in analysis.  Finally, another threat to validity was our lack of control over whether participants actually refrained from searching for answers over the Internet rather than asking our simulated virtual assistant. Participants could have used another device to search the web.  We did not notice any extended lapses in questions or work time from any participants, though, so we believe most participants followed our searching instructions correctly.

\vspace{-0.15cm}
\begin{figure}[ht]
\textbf{Project Name:} 2048\hfill \newline
\textbf{Bug Report:} The game board becomes unresponsive.\newline
	\label{fig:bug_example_code}
\verbatimfont{\small}
\begin{verbatim}
public GamePane(int size, BasePane basePane)
{
    this.size = size;
    this.basePane = basePane;
    setScore(0);
    this.tileSize = tileSizes[size];
    this.moveTime = 100 * 4 / size;

    setPrefSize(size * tileSize, size * tileSize);
    setLayoutX(175 - (size * tileSize) / 2);
    setLayoutY(175 - (size * tileSize) / 2);
    setStyle("-fx-background-color: #FFFFFF;");
    addTile();

    ... [Irrelevant code cut for paper space limitations]

    Thread focusField = new Thread(new Runnable()
    {
        @Override
        public void run()
        {
            while(!Thread.currentThread().isInterrupted()) {
                if(!isFocused()) {
                    try { Thread.sleep(100); }
                    catch (InterruptedException e) {
                        e.printStackTrace();
                    }
                requestFocus();
                }}}});
    focusField.setDaemon(true);
    focusField.start();
}
\end{verbatim}
\verbatimfont{\normalsize}
\caption{A description of a bug in the ``2048'' project with source code. Participants received full copies of the source code, however parts have been omitted for space limitations in this figure.}
\end{figure}

\vspace{-0.5cm}
\subsection{Data Collection}
\label{sec:user_simulation_data_collection}
We provided each participant with an Ubuntu 64-bit virtual machine.  We asked the participants to download the virtual machine ahead of the study. Inside the virtual machine, we provided Eclipse with all the buggy projects.  We also provided a screen recording program called SimpleScreenRecorder~\cite{SimpleScreenRecorder}.  We asked each participant to start the screen recording at the beginning of the study and leave the software running until the study had completed.  The participants then saved and sent the screen recording file to us.  We then collected the Skype transcript created by the study and removed all identifying information from the conversations.  Some participants also sent back the project files of the fixed bugs, but these files were not used in our analysis.           

\subsection{Bugs}
\label{sec:user_simulation_bugs}
The 20 Java bugs come from 17 different open source projects.  The project domains include a Pacman game, a calender application, and a PDF file merger.  We also selected bugs from commonly used Java libraries such as OpenCSV~\cite{OpenCSV} and Apache Commons IO~\cite{ApacheCommonsIO}.  We chose the bugs based on four criteria:
\begin{enumerate}
    \item The bugs had to be simple enough that they could be solved in a few hours, but complicated enough to take at least 20 minutes to solve.
    \item We had to be able to understand the bugs well enough to give meaningful answers to questions during the simulation.
    \item The user had to be able to reproduce the bug easily.
    \item The bugs had to be solvable without obscure domain knowledge.
\end{enumerate}

All of the bugs were previously solved, and we had the actual solutions on hand throughout each study.  However, we also discussed other solutions to the bugs before the user simulations.  This is because some of the bugs could be fixed in a variety of ways.  The bugs were presented individually and randomized for each study. An example of a bug given to the participants is as follows:

The bug in the source code above occurs when a user tries to make a move with the arrow keys. The source of the bug is the result of an incorrect fusion of a third party library used for graphics (JavaFX) and the structural design of the project. The project contains multiple panes which house buttons performing different types of actions. For the sake of simplicity, consider there to be only two panes; one that displays the board and is controlled by the arrow keys (the ``game pane''), and another that allows users to save and load games (the ``file pane''). Both of these panes are vying for focus, but for the game to be played, the ``game pane'' must always have focus.  To ensure this, the project's implementation spawns a deamon thread that almost constantly requests focus for the ``game pane.'' The bug comes from the fact that JavaFX only allows for one thread, called the ``event thread,'' to make changes to the UI. When creating the deamon thread, the developer uses the ``Thread'' type to request focus, which JavaFX interprets as modifying the UI. This causes an exception to be raised, and for the game to become unresponsive to the arrow keys.

One solution to this bug is to use JavaFX safe data types to perform the action of the deamon thread. During studies, participants were only provided with the buggy projects and the bug description. We (pretending to be Madeline), while aware of solutions, would in no form ``give'' a solution to the participants, but would only react to questions asked. Participants were incentivized to search the source project for the files containing bugs, as questions designed to tease solutions out of Madeline were met with vague and unhelpful responses (i.e. ``I am unsure''). A complete list of bugs can be found at our online appendix (see Section \ref{sec:reproducability}).

\vspace{-0.1cm}

\subsection{Experiences \& Lessons Learned}
\label{sec:user_simulation_experiences_lessons_learned}
In this section, we discuss our experiences and lessons learned while conducting the user simulation study.  We do this to provide some guidance to software engineering researchers who might do studies similar to ours in the future.
One of the biggest lessons we learned was to confirm that the virtual machine we provided worked on the participant's machine before the study started.  In roughly half of the studies, we found ourselves fixing problems on the participants' machines and spending, on average, an extra 20 minutes fixing the issues.  This was problematic, as the studies took up more time than originally anticipated, which threw off our original study schedule.  We also learned that additional information should be advertised (beyond the scope of the study) to allow for smooth experiments, such as experience with virtual machines or experience with the Eclipse IDE.

Another lesson learned was how to effectively schedule remote studies.  Many participants were unable to participate in the study until they returned home from their jobs in the evening.  Some had families and wanted to participate in the study even later, once their children were in bed. Many of our participants were in different time zones, there were days where we would schedule studies at 8 am, 1 pm, and 10 pm in our time zone.  We learned, over time, to hire participants overseas where their evening was our work day. 

\begin{figure*}[!t]
 \centering
 \includegraphics[width=0.95\textwidth]{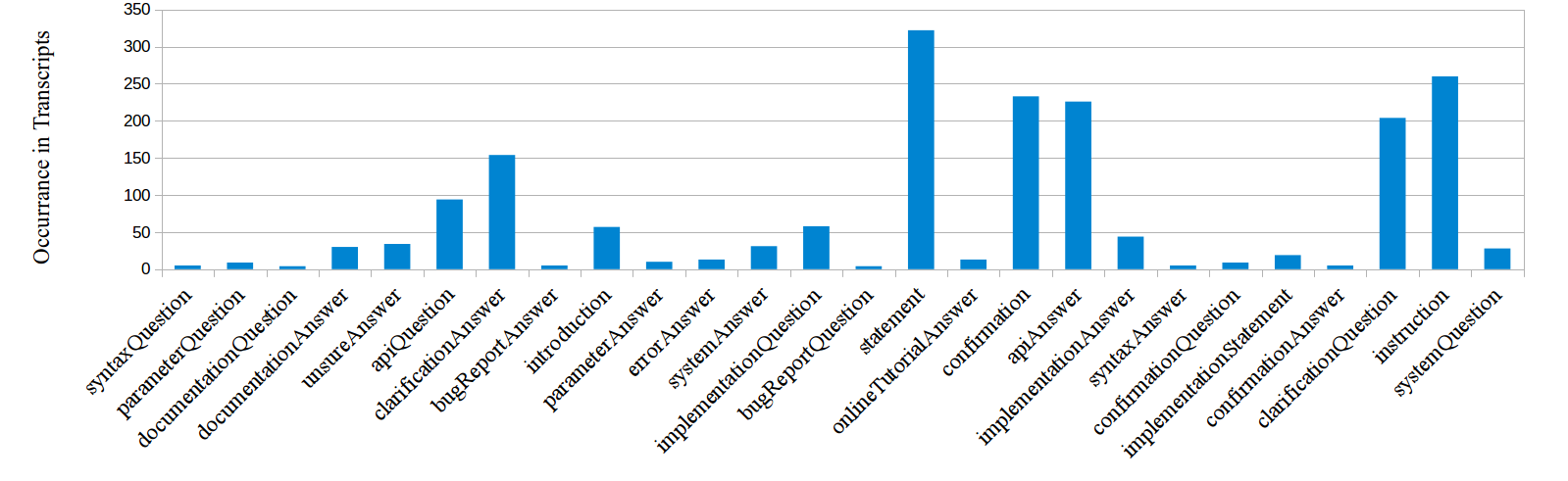}
 \caption{The annotation labels of all 30 transcripts and the occurrences for each label.  Each turn can have multiple labels (speech act type).}
 \label{fig:annotations}
\end{figure*}


\newcommand{\squeezeup}{\vspace{-2.5mm}}
\section{Annotations}
\label{sec:annotations}
In this section, we describe our process for annotating the speech acts from the data collected during the user simulation studies (see Section \ref{sec:user_simulation_data_collection}).  Essentially, our goal is to 1) determine what the speech acts are and 2) to determine what parts of the conversations are associated with those speech act types.  We also discuss our research questions, the rationale for asking them, and provide annotation examples.
\subsection{Research Questions}
\label{sec:annotations_research_questions}
The research objective of this section is to determine how programmers would use a virtual assistant to assist them in fixing a source code bug.  We seek to see what types of questions programmers would ask a virtual assistant and if those types of questions are consistent across multiple programmers.
\begin{description}
	\item[$RQ_{1}$] Do different programmers ask the virtual assistant similar questions for the same bugs?
	\vspace{0.1cm}
	\item[$RQ_{2}$] What types of questions did programmers ask during bug repair?
	\vspace{0.1cm}
	\item[$RQ_{3}$] What type of questions did programmers most frequently ask?
\end{description}
The rationale behind $RQ_1$ is that if programmers ask for help, and if they ask similar questions for the same bug, it is possible to create a speech acts classification system given training data. We group the questions to create labels for the training data in $RQ_2$. Finally, we investigate the most common uses of a potential virtual assistant in $RQ_3$ to advise future virtual assistant implementations.

\vspace{-0.17cm}
\subsection{Methodology}
\label{sec:annotations_methodology}
We used a manual open coding qualitative analysis process~\cite{berg2004methods} adapted from the social sciences to create labels for the conversations we collected.  (Though for the purposes of this paper, we follow Rastkar~\emph{et al.}~\cite{rastkar2014automatic} in referring to ``coding'' as ``annotating'' to prevent conceptual conflicts between sociological coding and computer coding.)  Qualitative annotation is becoming more common in software engineering literature~\cite{rodeghero2017detecting, murray2008summarizing, rastkar2010summarizing, dabbish2012social, hoda2010using}, and it is important to recognize that while standards and principles exist, the nature of qualitative analysis is that each annotation procedure is slightly different based on the needs and circumstances of the study.  In this paper, we followed the recommendations of Rieser and Lemon in a 2011 book on creating dialog systems from WoZ data~\cite{rieser2011reinforcement}, with one exception noted below.

A metaphor for open coding is unsupervised learning, in that the human annotators do not begin with a set of labels: our goal is to discover those labels from the data, and then assign them to the turns in our data.  Practically speaking, we did this in three rounds. The first round of annotation consisted of ``label creation'' where we created labels as we saw fit and did not have a pre-determined list to choose from. The second round consisted of ``label pruning'' where we decided what labels could be safely removed or merged. The second round became necessary the more progress was made in the first round, and was due to the complexity of compressing sometimes vague and complex English text down into its major concepts. The result of label pruning was a set of well defined and disjoint descriptions of English text describing our examples. The third and final stage of annotating involved re-examining the annotations but instead searching for spelling errors or other small mistakes. This round had the effect of ensuring labels were consistent and resolving labels that represented the same concept but used different terminology (i.e. synonyms), or were spelled incorrectly.

During any annotation process, and especially an open process in which we do not begin with labels, the bias of the human annotator becomes a major concern.  The degree of bias is known as the ``reliability'' of the data, and it is an extremely controversial research topic.  One possibility is to follow the lead of Carletta~\cite{Carletta:1996:AAC:230386.230390} in calculating Kappa agreement from multiple annotators, and only accepting agreement above a certain threshold; if agreement cannot be achieved, the argument goes, then more annotators are necessary.  While this is a common procedure, it is by no means universally accepted.  As Craggs and McGee Wood point out, ``one must decide for oneself, based on the intended use
of [an annotation] scheme, whether the observed level of agreement is sufficient''~\cite{Craggs:2005:EDD:1108994.1108995}.  Likewise, they ``suggest that if a coding scheme is to be used to generate data from which a system will learn to perform similar coding, then we should be `unwilling to rely on imperfect data'.''

At the same time, it is not an option to merely add more and more annotators until agreement is achieved.  There has long been a recognized split between expert and naive annotators~\cite{Carletta:1996:AAC:230386.230390, passonneau1993intention}.  It is not proper to allow naive annotators to have majority rule over the experts.  To be an expert annotator in our study, a person would need to have 1) knowledge of the bugs solved in our study so they can understand the conversations, and 2) not been a participant in the study.  Only the first and second authors were both qualified and available (manual annotation is weeks of effort).

Rieser and Lemon faced a similar situation, and solved it by having a discussion between two annotators for all disagreements, followed by independent decision-making and calculation of Kappa (page 110 of \cite{rieser2011reinforcement}).  We differ from this procedure in that we consider our situation to be more ``unwilling to rely on imperfect data'' due to the fact that our research questions in Section~\ref{sec:annotations_research_questions} and our prediction training in Section~\ref{sec:predictapproach} could be affected by errors.  Therefore, for this paper, we had two experts annotate all data and solve every disagreement through discussion as disagreements occurred, followed by mutual decision-making, resulting in one set of annotations.  While this mutual process makes it impossible to calculate a reliability metric, we felt it was more important to maximize correctness of the annotations.


\vspace{-0.1cm}
\section{Annotations Results}
\label{sec:annotations_results}
In this section, we present the results from our annotation process.  We also provide annotation examples following the results.  We note that the programmers asked on average 12.8 questions throughout the two hour user simulation study.  A select few did not ask more than three, however, these participants were outliers.  The highest number of questions asked during a user simulation study was 54 and the lowest number of questions asked during a study was 3.


\vspace{-0.1cm}
\subsection{$RQ_{1}$: Programmers asking similar questions}
We found that programmers asked similar questions to one another.  Of all the questions asked by the programmers, the ones that were consistent across the majority of participants included confirmationQuestions, clarificationQuestions, and apiQuestions.  Of these three types of questions, clarificationQuestion was asked the most by all programmers.  It was asked a total of 204 times, which comprised 53.1\% of all questions asked by programmers.  There were various types of clarification questions asked.  Some of the clarification questions included questions about what the bug report said, what questions Madeline could and could not answer, and clarifying answers from Madeline.  The participants also asked clarification questions to confirm their understanding of the task that they were to complete for the study.  


\vspace{-0.1cm}
\subsection{$RQ_{2}$: Types of questions being asked}
\label{sec:annotations_types_of_questions_being_asked}
We found that programmers asked a variety of questions that ranged from system type questions to API and documentation types. An example of an API question is:

\begin{verbatim}
   "What methods are in eventyhandler(?)"
\end{verbatim}
We also found many programmers asked implementation questions:
\begin{verbatim}
   "What are valid values for the int direction
    in PacPlayer.java?"
\end{verbatim}
After finishing the annotation process, we were able to narrow down the question annotation types into 10 categories.  The categories are: syntax, parameter, documentation, API, clarification, implementation, bug report, confirmation, clarification, and system questions. Figure~\ref{fig:annotations} lists the number of occurrences for each of the speech act types.  In Section \ref{sec:annotation_results_annotation_examples} we go into detail with a short example of an annotated conversation.    We also provide all of the annotations on our online appendix (see Section \ref{sec:reproducability}).      

\vspace{-0.1cm}
\subsection{$RQ_{3}$: Most frequent questions being asked}
We found programmers asked a few questions significantly more than others. In Figure~\ref{fig:annotations}, the speech act type ``statement'' has the most occurrences.  We would like to point out that there was another, more popular type of question; the ``setup'' speech act. Since this speech act type is not relevant to the study itself, this speech act type was removed from our corpus.  ``clarificationQuestion'' has the highest occurrence out of any question type.  This label appeared 204 times throughout all 30 transcripts.  Many of the participants asked clarification questions on the bugs and on the responses Madeline gave. Madeline asked clarification questions as well when we needed more information from a participant to answer a question.  Sometimes the participants would ask questions that needed more detail so that Madeline could answer the question.  The second highest occurrence annotation label for a question type was   ``APIquestion.'' This label occurred 94 times in the transcripts.  This makes sense as programmers were not allowed to use the internet during the bug repair task and were unfamiliar with the given source code.   

\vspace{-0.2cm}
\subsection{Annotation Examples}
\label{sec:annotation_results_annotation_examples}
We annotated over two thousand speech acts during the annotation process.  To further explain the previous sections, we provide an example of one of the annotations.  Throughout the data, participants asked API questions, documentation, and implementation questions.  Below is a section of a developer conversation.  This section of the conversation includes implementation questions and clarification questions.  At the end of each speech act, there is the annotation label for that speech act.  The annotation is in bold text and is in brackets.  The speech acts begin with ``P'' or ``M'' denoting the speaker as a ``participant'' or ``Madeline - Virtual Assistant'' respectively.\break
\vspace{-0.2cm}
\begin{tcolorbox}
P: So the bug is that the PacPlayer does not face right when the key is released, but it is supposed to? \hfill
\break
\textbf{[clarificationQuestion]}  \hfill
\vspace{-0.2cm}
\break
\break
M: Yes. He also disappears. \textbf{[clarificationAnswer]} \hfill
\vspace{-0.2cm}
\break
\break
P: Does he disappear because the alive bool is 
set to false at the wrong time \textbf{[implementationQuestion]}  \hfill
\vspace{-0.2cm}
\break
\break
M: I am unsure \textbf{[unsureAnswer]} \hfill
\end{tcolorbox}

Throughout the annotation process, we found similar results to the previous example. However, we found programmers asked varying amounts of questions throughout the bug repair task.  This was evident once deep into the annotation process.  It appeared that the more senior a participant was, the less the participant asked for help from the virtual assistant.  There are three interpretations we derive from these observations.  First, the programmers possibly did not want to ask for help and instead wanted to solve the bug without help.  Second, it is possible that the programmers did not feel comfortable asking questions. Finally, the programmers may have assumed that there was no automated virtual assistant and, therefore, did not ask questions.

We found that programmers often made a statement before asking a question.  It appeared the participants were explaining their thought process before asking a question.  This occurred about 20\% of the time in the user simulation studies. An example of this is:


\begin{tcolorbox}
participant: first I tried ``sudo apt-get install default-jre'' \hfill
\vspace{-0.2cm}
\break
\break
participant: it told me it depends on default-jre-headless and openjdk-7-jre \hfill
\vspace{-0.2cm}
\break
\break
participant: is it possible to set a command line argument for start up of the program?  \hfill
\end{tcolorbox}

Here, the participant makes multiple statements before asking Madeline a question.  We did not ask participants to ``think aloud'' during this study.  However, we observed this phenomenon throughout the user simulations and annotation process. 


\section{Predicting Speech Act Type}
\label{sec:predictapproach}

Our approach for predicting the speech act type is, essentially, a text classifier based on Logistic Regression.  Recall the use case that we envision in Section~\ref{sec:problem}: a virtual assistant receives a message, and needs to classify that message into one of several categories, so that it can respond appropriately.  Our idea is to train a prediction model, then use that prediction model to classify incoming messages.

\vspace{-0.1cm}
\subsection{Labeled Training Data}

Supervised machine learning algorithms depend on labeled training data.  We use the labels from Section~\ref{sec:annotations_types_of_questions_being_asked}.  In that section, we manually annotated every turn in every conversation as belonging to one of the speech act types we identified.  In this section, we use that data (however, only turns from the participants' side of the conversation, not ``Madeline's'', to match the use case of the virtual agent classifying incoming messages) to train a classifier to annotate the turns automatically.  Note that this is a multi-label classification problem, because an ``example'' consists of a turn annotated with a list of all the speech act types to which that turn belongs.  Each speech act turn type is a label, so each turn may belong to many labels.

\vspace{-0.1cm}
\subsection{Attributes}
\label{sec:attributes}

We use two types of attributes.  First, we treat the problem as text classification, where each word is an attribute.  We calculate the attributes as a binary ``bag of words''.  For each example, the set of attributes includes either a one or zero for each word, depending on if that word occurs in the text of the turn or not.  Recent industry-track papers~\cite{7816508, CruzRE17} came to the conclusion that to maximize potential industrial impact, researchers should prioritize simplicity, and only move to more complex approaches when absolutely necessary.  We stuck to binary bag of words for this reason.  We also did not do stop word removal or stemming.  We defer word count normalization (e.g. TF/IDF), NLP-based solutions, advanced preprocessing techniques, etc., to our future work.  As we will explain in Section~\ref{sec:predictevalresults}, the simple approach already achieves reasonable performance.

Second, we used three shallow features identified by related literature~\cite{Rodeghero:2017:DUS:3097368.3097375, rastkar2014automatic, murray2008summarizing}.  This related literature actually identifies over twenty shallow features that complement or replace text classification, but many of these are not applicable in our context.  For example, many rely on computing entropy over a whole conversation after the fact.  That is not possible in our context because we can only know incoming message and previous messages, not future messages.  The three features we used are: \texttt{slen}, the number of words in the message normalized over all previous messages, \texttt{wc}, the number of words not normalized, and \texttt{ppau}, the number of seconds between the message and the previous message.

\vspace{-0.1cm}
\subsection{SMOTE}
\label{sec:predictapproach_smote}
We use SMOTE~\cite{chawla2002smote} to overcome the problem of unbalanced data.  Some of the user speech acts we identified only have a few examples (e.g. we only found eight examples for the \texttt{parameterQuestion} type).  That presents a problem because the learning process will inevitably classify no turns in that type, and still seem to achieve very high accuracy.  SMOTE works by synthesizing examples in small classes from the known examples in those classes.  The result is that the small classes are filled with synthesized examples until the data are balanced.  SMOTE has been widely used to resolve situations of unbalanced data generally as well as conversational analysis~\cite{Rodeghero:2017:DUS:3097368.3097375}.  In pilot studies, we compared SMOTE to duplicative oversampling and observed slight performance improvements using SMOTE.  We used SMOTE only on the training data, to avoid biasing the testing set.

\vspace{-0.1cm}
\subsection{Prediction Models}
\label{sec:predictapproach_prediction_models}
We trained a multi-label prediction model using the \emph{binary relevance}~\cite{read2011classifier} procedure.  The procedure is to create one binary classifier for every class.  We used the Logistic Regression (LR) algorithm~\cite{hosmer2004applied} to create each classifier.  We also tested Naive Bayes and Support Vector Machines in pilot studies -- LR had superior performance to Naive Bayes, and the difference between LR and SVM was so slight as to not be worth the much longer training time for SVM (eight hours versus four minutes).  Note that while we built a multi-label prediction model, we calculated SMOTE using a multi-class structure.  That is, we ran SMOTE once for each category, then trained each classifier, then combined the classifiers with the binary relevance procedure.  In theory it is possible to run SMOTE in a multi-label configuration, by executing SMOTE on every combination of labels.  However, this would necessitate $n^n$ runs of SMOTE (for $n$ categories), which would be far more expensive.

We also performed parameter tuning for Logistic Regression across twelve parameters and Naive Bayes across four parameters.  Parameter tuning has been recommended generally when working with SE data~\cite{Binkley:2014:ULS:2597008.2597150}.  Due to space requirements, we direct readers to our online appendix and reproducibility package for complete details (see Section~\ref{sec:reproducability}).


\subsection{Implementation Details}
\label{sec:predictapproach_implementation_details}
We used the toolkit \texttt{scikit-learn}~\cite{pedregosa2011scikit, scikit-learn} to implement our classifiers and SMOTE (\texttt{imblearn.over\_sampling.SMOTE})~\cite{JMLR:v18:16-365}.\\
We implemented the shallow attribute calculators ourselves, using related work as a guide~\cite{Rodeghero:2017:DUS:3097368.3097375}.  The hardware was an HP Z640 workstation with an E1630v3 CPU and 64GB of memory.  For total clarity, we make all implementation scripts and datasets available via our online appendix (see Section~\ref{sec:reproducability}).

\vspace{-0.1cm}
\section{Evaluation of Predictions}

This section describes our evaluation of the prediction models we create.  Essentially, we use a 5-fold cross validation procedure to test the quality of the predictions, as well as explore where the predictions are most accurate.

\vspace{-0.1cm}
\subsection{Research Questions}

Our research objective is to determine what level of performance we can expect from the prediction models, as well as to understand which speech acts are ``easiest'' to detect.

\begin{description}
	\item[$RQ_{4}$] What is the performance of our prediction models, overall in the multi-label configuration, according to the metrics described in Section~\ref{sec:metrics}?
	\vspace{0.1cm}
	\item[$RQ_{5}$] For which speech acts do the prediction models have the highest performance?
	\vspace{0.1cm}
	\item[$RQ_{6}$] Which attributes are the most informative?
\end{description}

The rationale behind RQ$_4$ lies in the application we intend in Section~\ref{sec:problem}: the performance of a virtual assistant will be limited by its ability to detect what type of speech act to which an incoming message belongs.  While we do not expect perfect performance, we need to at least have an understanding of how much inaccuracy may stem from the detection process.  The rationale behind RQ$_5$ is similar.  Some speech acts are bound to be easier to detect than others.  It is helpful to know which speech acts about which we may be confident, or others where we are less sure.  In practice, it may be necessary to return a message to the user indicating that the virtual assistant is unsure what the user intends, and ask the user to clarify.  RQ$_6$ is useful because the presence of some attributes may indicate high confidence, while others may indicate low confidence.


\subsection{Methodology}

In general, we follow a 5-fold cross validation study design.  In a standard $n$-fold design for evaluating classifiers, $1/n$ examples are set aside as a testing set, while the remaining $(n-1)/n$ examples are used for training.  The evaluation is conducted $n$ times, once for each $n^{th}$ selection of the examples as a testing set.  Then, the evaluation metrics are calculated for each ``fold'' and averaged.  We chose 5 as a value for $n$ because it ensured that our testing set would not be too small (as it might have been with a 10-fold design), while still maintaining multiple folds that could be averaged.

The selection of a testing set is a non-trivial exercise in a multi-label dataset, in contrast to a single-label one.  In a single-label dataset, it is usually sufficient to randomly selected $1/n$ of the examples for the testing set.  But in our multi-label dataset, we need to ensure that the testing set represents the same distribution of labels as the overall dataset.  With only five folds, it is conceivable that a random selection would give too much weight to one label, and this overweighted selection would not be ``averaged out'' over a large number of folds.  Therefore, we sampled each label in proportion to the number of examples in that label, and confirmed that the distribution of the labels over the testing set was as close as possible to the distribution of labels over the entire dataset.

After we separated the testing and training data, we ran SMOTE on the training data only.  If we had executed SMOTE on the entire dataset, then divided the data into testing/training groups, we would have contaminated the testing set with information from the training set.  SMOTE synthesizes examples based on the examples it is given.  If we had run SMOTE on the entire dataset, we would have created synthesized examples based on real examples that ended up in testing set.  Therefore, we only ran SMOTE on the training set.  This did increase the execution cost of our experiment slightly, since we needed to execute SMOTE five times (once for each fold, after we separated the testing set from the training set).

Note also that this methodology is conservative -- it only uses real examples for the testing set.  We use the results from this conservative approach to answer RQ$_4$ and RQ$_5$, to avoid presenting a biased result.  We also use these results to calculate other metrics (see the next section) to answer RQ$_6$.  

\subsection{Metrics}
\label{sec:metrics}

We report the metrics precision, recall, F-measure, and support to answer RQ$_4$ and RQ$_5$  These are standard metrics for evaluating classifiers and have been covered extensively elsewhere~\cite{batista2004study, caruana2006empirical}; for space we do not discuss their details here.  We calculate these metrics for each speech act type (i.e., each label) for RQ$_2$, and combine the results for each speech act type to answer RQ$_4$.  We combine the precision and recall values for each speech act type with a weighted average, where the weights are based on the support for each speech act type.  The reason is so that the combined values reflect the size of each label.  A simple average, without the weights, would be biased by labels that only have a few examples in the testing set.

For RQ$_6$, we calculate F-score~\cite{Rijsbergen:fscore} for the attributes.  F-score (distinguished from F-measure, the harmonic mean of precision and recall) is typically used for feature selection, to indicate which features are the most informative.

\subsection{Threats to Validity}

Like all experiments, our study carries threats to validity.  The main sources of threats to validity include: the participants in the user simulations, the bugs we asked the users to repair, and the influences of the technology used by the participants (e.g., the IDE) on the questions they asked.  Also, it is possible that there are errors in our manual annotation process, or in our selection of categories.  While we try to mitigate these risks by following accepted data collection and annotation procedures, and by including a relatively large number of participants (30) and different bugs, the threat remains that changes in these variables could affect the performance of our classifiers.  As an additional guard against these risks, we release all data via an online appendix for community scrutiny (see Section~\ref{sec:reproducability}).

\section{Prediction Eval. Results}
\label{sec:predictEvalResults}

This section discusses our answers to RQ$_4$-RQ$_6$, including our supporting data and rationale.

\begin{table}[ht]
	\centering
	\caption{Performance metrics calculated for each speech act type (some speech act types have been abbreviated).  Recall that the averages are a weighted average based on the support for each speech type, see Section~\ref{sec:metrics}.}
	\vspace{-0.3cm}
	\label{tab:performance}
	\begin{tabular}{lllll}
		& precision & recall & f-measure & support \\
		apiAnswer             & 0.93      & 0.76   & 0.83    & 24.6    \\
		apiQuestion           & 0.81      & 0.66   & 0.71    & 17.2    \\
		clarifAnswer          & 0.13      & 0.07   & 0.09    & 6.0     \\
		clarifQuestion        & 0.59      & 0.41   & 0.48    & 32.6    \\
		confirmation          & 0.88      & 0.8    & 0.83    & 27.0    \\
		docAnswer             & 0.25      & 0.2    & 0.22    & 3.2     \\
		implQuestion          & 0.52      & 0.21   & 0.28    & 10.6    \\
		implStatement         & 0.0       & 0.0    & 0.0     & 3.0     \\
		introduction          & 0.76      & 0.6    & 0.63    & 4.0     \\
		stmnt                 & 0.69      & 0.4    & 0.51    & 49.8    \\
		systemQuestion        & 0.37      & 0.22   & 0.27    & 4.8     \\
		\hline
		avg / total           & 0.69      & 0.5    & 0.57    & 16.62  
	\end{tabular}
\end{table}

\vspace{-0.2cm}
\subsection{RQ$_4$: Overall Performance}
\label{sec:predictevalresults}

\begin{table*}[ht]
	\centering
	\caption{The top 10 most-informative features for each speech act type, calculated by f-score.  Most features are words, but features with the suffix \texttt{\_sf} are shallow features (see Section~\ref{sec:attributes}).  See Section~\ref{sec:attributeeffects} for a deeper discussion of this table.}
	\label{tab:words}
	\vspace{-0.2cm}
	{\footnotesize
		\begin{tabular}{lllllllllll}
			& 10        & 9           & 8            & 7                    & 6                      & 5                       & 4                & 3           & 2                      & 1           \\
			apiAnswer   & node  & if    & onfinished    & keyframes & constructor   & values    & time  & timeline  & keyvalue  & keyframe   \\
            apiQuestion & size  & method    & have  & how   & pane  & class & object    & an    & does  & what \\
            clarifAnswer & compilation   & configurations    & word  & trigger   & supply    & green & appear    & box   & clicking  & bottom \\
            clarifQuestion   & need  & the   & or    & other & wc\_sf    & you   & this  & fix   & prime bug \\
            confirmation    & of    & yes   & is    & to    & thanks    & slen\_sf  & the   & thank & wc\_sf    & ok   \\
            documentAnswer & byte  & marks & later & reading   & bytes & joptionpane   & external  & input & audio & stream  \\
            implementQuestion  & face  & why   & mark  & eratosthenes  & occurs    & gets  & arraycopy & reason    & button    & clicked   \\
            implementStatement & signature & widget    & funtion   & hidden    & drawing   & waitfor   & throwing  & paint & timeout   & jcomponent   \\
            introduction    & supervised    & programmers   & today & hello & human & am    & start & hi    & study & ready   \\
            statement   & seems & what  & slen\_sf  & looks & fixed & but   & works & was   & think & it    \\
            systemQuestion  & password  & there & permitted & lang  & running   & way   & programs  & kill  & eclipse   & how  
		\end{tabular}
	}
\end{table*}

The weighted average precision of from our classifiers was 69\%, while the weighted average recall was 50\%, as reported in Table~\ref{tab:performance}.  Thus for an arbitrary incoming message, we can expect this classifier to correctly identify the speech act type of that message 69\% of the time, while identifying 50\% of the speech acts types to which the message belongs.  If the classifier claims that a message of a particular type, we can estimate that that claim will be correct roughly 2/3rds of the time.  We acknowledge that we cannot evaluate whether these improve over an earlier approach, given that we are not aware of an earlier technique for identifying speech acts on our data.  
 Nevertheless, we find these results to be an encouraging starting point for building a virtual assistant, in light of the somewhat bare bones text classification strategy we used (binary bag-of-words, see Section~\ref{sec:predictapproach}).  A promising area of future work, in our view, is to adapt more advanced classification techniques.

\vspace{-0.1cm}
\subsection{RQ$_5$: Speech Act Type Variations}

The performance of our classifiers varied considerably across different speech act types.  At the high end, precision was over 90\% and recall over 75\%.  At the low end, precision and recall dipped below around 10\%.  This observation is important because it means that for some speech act types, a virtual assistant can be highly confident that the prediction is correct.  As a practical matter, a virtual assistant may request the user to repeat a message in different words, or ask for other followup information, if the classifier is not able to place the message into a speech act type with sufficient confidence.  This observation is also important from an academic viewpoint, because it means that programmers use different types of language to make different types of messages.  In some cases, programmers consistently use the same language (which is what the classifier uses to make good predictions).  In other cases, programmers use much more different language -- it makes the prediction process more challenging, but also raises academic questions about what is different about the language, which is an area of future work.  We begin to explore this in RQ$_6$.

\vspace{-0.1cm}
\subsection{RQ$_6$: Attribute Effects}
\label{sec:attributeeffects}

Table~\ref{tab:words} shows the top-10 most informative features for each speech act type.  We make two observations from this data: First, the shallow features are far more useful for some speech act types than others.  For example, confirmation actions are likely to be short messages, so the word count metric (\texttt{wc\_sf}) is informative in this case.  This observation is useful because shallow features are easy to compute, so areas where they are informative can be predicted with reasonable accuracy at low cost.  Second, many of the words are general enough that they are likely to be generalizable beyond the set of bugs we chose, even though others are specific to particular domains.  For example, the speech act \texttt{implementationStatement} is informed by words like ``function'' and ``signature'', which are likely to be true across many programming conversations.  But the most informative feature for that action is ``jcomponent'', which is a word specific to Java and perhaps the domain of programs we study.  It is not likely to appear in every domain.  Therefore, one possible mediation is to use placeholder features that count the number of e.g. domain-specific programming words used in a message.  Also, we note again that we used the binary bag-of-words model, which separates the words from their contexts.  An area of future work is in NLP-based recognition such as phrases or n-grams.

\vspace{-0.2cm}
\section{Conclusion}
\label{sec:conclusion}
Our paper makes three contributions to software engineering literature.  First, we contribute 30 software engineering conversations with professional developers. Second, we created a system of classification for developer speech acts.  We manually detect and classify relevant speech acts in order to contribute to the understanding of developer question/answer conversations.  We also provide this annotation classification system on our online appendix for future researchers to use.  Third, we lay the foundation for a virtual assistant by building an automatic speech act classification system.

\vspace{-0.2cm}
\section{Reproducibility}
\label{sec:reproducability}
We have made our raw data, annotations, model, and source code available via an online appendix (\url{https://tinyurl.com/yadfpojd}) for the research community to reproduce or use. 

\vspace{-0.2cm}
\section{Acknowledgements}
\textit{We  thank  and  acknowledge  the 30 professional developers who participated in this research study.  This work is supported in part by the NSF CCF-1452959, CCF-1717607, and CNS-1510329 grants. Any opinions, findings, and conclusions expressed herein are the authors' and do not necessarily reflect those of the sponsors.}.


\bibliographystyle{ACM-Reference-Format}
\bibliography{main} 

\end{document}